\documentclass[conference]{IEEEtran}
\IEEEoverridecommandlockouts
\usepackage{cite}
\usepackage{amsmath,amssymb,amsfonts}
\usepackage{algorithmic}
\usepackage{graphicx}
\usepackage{textcomp}
\usepackage{xcolor}
\usepackage{hyperref}
\usepackage{cleveref}
\usepackage{booktabs}
\usepackage{makecell}

\def\BibTeX{{\rm B\kern-.05em{\sc i\kern-.025em b}\kern-.08em
    T\kern-.1667em\lower.7ex\hbox{E}\kern-.125emX}}

\newcolumntype{P}[1]{>{\raggedright\arraybackslash}p{#1}}

\begin{document}

\title{SoK: Behind the Accuracy of Complex Human Activity Recognition Using Deep Learning}

\author{
\IEEEauthorblockN{Duc-Anh Nguyen}
\IEEEauthorblockA{\textit{School of Computer Science} \\
\textit{University College Dublin}\\
Dublin, Ireland \\
0000-0001-8613-6854\\
duc-anh.nguyen@ucdconnect.ie}

\and

\IEEEauthorblockN{Nhien-An Le-Khac}
\IEEEauthorblockA{\textit{School of Computer Science} \\
\textit{University College Dublin}\\
Dublin, Ireland \\
0000-0003-4373-2212\\
an.lekhac@ucd.ie}
}

\maketitle

\begin{abstract}
Human Activity Recognition (HAR) is a well-studied field with research dating back to the 1980s. Over time, HAR technologies have evolved significantly from manual feature extraction, rule-based algorithms, and simple machine learning models to powerful deep learning models, from one sensor type to a diverse array of sensing modalities. The scope has also expanded from recognising a limited set of activities to encompassing a larger variety of both simple and complex activities. However, there still exist many challenges that hinder advancement in complex activity recognition using modern deep learning methods. In this paper, we comprehensively systematise factors leading to inaccuracy in complex HAR, such as data variety and model capacity. Among many sensor types, we give more attention to wearable and camera due to their prevalence. Through this Systematisation of Knowledge (SoK) paper, readers can gain a solid understanding of the development history and existing challenges of HAR, different categorisations of activities, obstacles in deep learning-based complex HAR that impact accuracy, and potential research directions.
\end{abstract}


\section{Introduction} \label{sec:intro}
Human Activity Recognition (HAR) is the process of automatically determining human activities from sensor data. HAR has a wide range of applications, namely, health monitoring, surveillance, sport analysis, human-machine interaction, and signal language translation.
Depending on applications, activities of interest can vary, from simple and repetitive like walking, specific like a particular hand gesture, to complex like falling and cooking. While heuristic and shallow machine learning may recognise simple activities, deep learning offers the capability to discern intricate patterns in complex activities.
Also, the choice of sensor types varies by numerous factors like costs of operation, user acceptance, privacy concerns, and activities' characteristics. Cameras are suitable to observe one or many people within a certain area, other ambient sensors (e.g. infrared sensor, radar) are suited for indoor usage, and wearables are convenient for individual usage.

Accuracy of HAR depends on many factors, namely sensor device, data quality, recognition method, and activities of interest. Simple activities are almost always easier to recognise than complex ones, despite that it is not the only factor. Also, while the other three factors can be controlled, activities of interest are totally determined by application requirements and users. Therefore, attention should be given to complex activity recognition to address this challenge. Though there have been many published works in the field of HAR, most methods and public datasets only focused on simple activity data \cite{Ramanujam2021}. 

To solve a problem, it is crucial to comprehend the nature of the issue and the intended goal. This paper presents a systematisation of knowledge on factors contributing to inaccuracy in complex HAR, an area that has not been extensively studied. We also attend to fall detection as falling is one of the most studied complex activities.
Our contributions are as follows:
\begin{itemize}
    \item We provide a chronological overview of HAR research, highlighting the evolving research landscape and the remaining challenges of modern HAR systems.
    \item We systematise the factors that frequently contribute to inaccuracy in complex HAR using deep learning, both in model testing and real-world usage. These are crucial for achieving robust complex HAR.
    \item We discuss the potential methods as well as promising research ideas for tackling the above problems, featuring challenges and opportunities.
\end{itemize}

In \Cref{sec: evolution}, we review HAR's evolution from the 1980s to the present and summarise the remaining problems. \Cref{sec: categorisation} defines categorisations of human activity to avoid confusion between complex activity and other terms like composite activity. \Cref{sec: factors} systematises the factors affecting the accuracy of deep learning-based complex HAR. Finally, \Cref{sec: towards} discusses potential directions for further research.

\section{Evolution of HAR Technologies}\label{sec: evolution}

\subsection{From shallow to deep learning}
\textit{During the 1980s and 1990s}, HAR research was predominantly done using camera. The common approach involved extracting features from images for subsequent activity recognition. Methods requiring human body joint coordinates had to rely on markers attached to the subject's body. Other feature extraction techniques involved optical flow, motion images, and background subtraction. Their drawbacks were the assumptions made, such as static background, and user's specific positions and motions. For activity recognition, template matching approaches transform an image sequence into a static form, whereas state-space approaches analyse the whole sequence. These approaches employed machine learning models like kNN, Decision Tree, Regression, and Hidden Markov Model (HMM) \cite{Aggarwal1999}. Some studies applied neural networks like Multi-layer Perceptron (MLP) on image sequence \cite{Wang2003}.

\textit{In the 2000s}, methods using human silhouette dominated. Also, point trajectories, background subtracted blobs and shapes, and filter responses were proposed. Efforts were made to track body parts from normal images instead of using body markers. Methods dealing with camera in motion were also explored. Dimension reduction methods like Principal Component Analysis (PCA) were applied after feature extraction. Classification methods became more diverse, some worked with the whole human body, others looked at individual body parts. Researchers also attempted unsupervised and semi-supervised approaches. Methods also varied in how they processed input features. Some considered separate frames before combining them, others treated videos as 3D pixel volumes, and parametric time-series approaches imposing models on temporal motion dynamics were also employed \cite{Moeslund2006}. Some commonly used models were HMM, kNN, Gaussian classifier, and Bayes networks. Despite the improvements, many problems were left open, namely real-world environment, intra-activity variability, image resolution, and occlusion \cite{Turaga2008}.

\textit{The 2000s} also witnessed a rise in HAR research using wearable sensors on different body parts, such as waist and arm, to capture motions in 3D space (e.g. acceleration, rotation). Temporal segmentation was investigated to divide time series into windows of individual activities before classification using sliding window, bottom-up, and top-down approaches. Some commonly used features were time domain features (mean, variance, interquartile range, mean absolute deviation, entropy, root mean square), frequency domain features (spectral energy, spectral entropy, spectral centroid), and heuristic features (signal magnitude area, signal vector magnitude, inter-axis correlation). Features could be selected using SVM, K-means, and forward-backward sequential search; or transformed using PCA, Independent Component Analysis, and Local Discriminant Analysis. Classification models were threshold-based, Decision Tree, kNN, Naive Bayes, Support Vector Machine (SVM), HMM, Gaussian Mixture Model (GMM), and neural network \cite{Avci2010}. Unsupervised and semi-supervised approaches were studied for wearables as well \cite{Lara2013}.

\textit{In the early 2010s}, deep learning was adopted for vision-based HAR. Convolutional Neural Networks (CNN) became the first choice of feature extractor for many tasks. Both 2D and 3D CNNs were used, and 3D CNNs were preferable for spatial-temporal data like videos. As the spatial dimension (image size) was fixed, attention was drawn to answering how temporal data could be fed into CNNs. Recurrent Neural Networks (RNN) and Long-Short Term Memory (LSTM) were also used to learn temporal information \cite{Herath2017}. For sensing devices, advances in depth camera have made 3D human pose tracking easier. Also, the fusion between depth and inertial sensor was explored at 3 levels: data, feature, and decision \cite{Chen2017}. As for wearables, researchers continued to investigate supervised and unsupervised machine learning models like SVM, Random Forest, kNN, HMM, K-means, and GMM \cite{Attal2015}. Research was also done for the fusion of multiple sensors worn on the subject's body with algorithms like Kalman filter, complementary filter, integration, and vector observation \cite{Nava2016}.

\textit{In the late 2010s}, deep learning has become popular for HAR research using both sensor types and various fusion approaches. Many deep networks were built based on architectures like CNN, RNN, LSTM, Graph Neural Network (GNN), and multi-stream networks \cite{PranjalKumar2023}. Many learning algorithms were investigated: ensemble combines multiple models, unsupervised learning (Autoencoders, Deep Belief Network) and semi-supervised learning (active learning, synthesised data, multimodal co-learning) mitigate labelled data scarcity, and transfer learning reduces the effects of domain discrepancy. Data segmentation was handled using segmentation models and multi-label models. Other aspects of HAR like user privacy and computational cost have gained more attention as well \cite{Chen2021}. Also, researchers have integrated more sensing modalities (e.g. GPS, WiFi, Bluetooth, heart rate, audio, and passive infrared sensor), widening the application range \cite{Demrozi2020}.

\textit{From 2020 to present}, deep learning models have been proposed based on Transformer and other prior architectures. The community has also been exploring semi-supervised and self-supervised methods to make the most of unlabelled data. Examples of some investigated learning algorithms are Contrastive Learning, Generative Adversarial Network (GAN), and Stacked Denoising Autoencoder (SDA) \cite{PranjalKumar2023,Sun2023}.

\subsection{Remaining problems}
HAR technologies have evolved significantly over several decades. While early research explored many challenges and approaches like complex HAR, sensor fusion, and semi-supervised learning, the advent of deep learning has raised the bar for HAR performance and versatility. For example, activities recognisable by earlier methods are now tested in much more complex and diverse scenarios. As HAR technologies continue to advance, they encounter critical challenges that shape their efficacy and development.

Current HAR systems face difficulty in recognising complex activities and activities with unknown or evolving patterns, such as illegal behaviours in surveillance applications. The collection of large-scale labelled HAR datasets is a major hurdle as it involves human subjects. The lack of context in activity data, inaccurate labels, and ambiguity in event occurrence timing hinder accurate recognition. Also, class imbalance, inter-activity variability, and differences between real-world and experimental data complicate the task. For deep learning, optimising hyperparameters is an important task, requiring computationally expensive evaluations. The lack of reproducibility and transparency of research papers, and the lack of standardisation in testing measures and benchmarks hinder comparisons of different approaches \cite{Qiu2022,Islam2022,PranjalKumar2023}.

\section{Activity Categorisations in HAR Context}\label{sec: categorisation}
Before getting into the causes of inaccuracy in complex HAR, we present three activity categorisations under the HAR context to avoid any confusion and to facilitate a systematic and insightful analysis in \Cref{sec: factors}.

\subsection{Categorisation by complexity}
Complex activity is the target of this paper. Though it is hard to draw a clear line between complex and simple activities, we categorise activity based on complexity as follows:
\begin{itemize}
    \item \textbf{A simple activity} consists of an action or a series of repeated actions with clear patterns. There is little variation among individuals when performing this activity. E.g. chopping food, walking, drawing a circle, tripping.
    \item \textbf{A complex activity} consists of an action or a series of dissimilar actions. It is performed inconsistently among individuals or even among repetitions of the same individual. E.g. cooking, moving a chair, painting, falling. Though falling is a short and noticeable movement, we categorise it as being complex because it varies immensely among instances.
\end{itemize}

\subsection{Categorisation by structure}
The term \textit{complex activity} may be confused with \textit{composite activity} or \textit{concurrent activities} because they have a high chance but are not guaranteed to be complex. Activities can be categorised by their structure as below:
\begin{itemize}
    \item \textbf{A single activity} or an individual activity is a clearly defined action that can be recognised independently. E.g. falling, walking, watching film.
    \item \textbf{A composite activity} is a sequence of interconnected single activities. It also includes transitions between individual activities. E.g. cooking includes many steps, an exercise includes multiple sub-activities.
    \item \textbf{Concurrent activities} are activities performed simultaneously. They often have some connection with each other but could also be performed separately. E.g. eating popcorn while watching film, talking while walking.
\end{itemize}

\subsection{Categorisation by interaction}
This categorisation is less known than the others. Activities can be divided by interaction as below:
\begin{itemize}
    \item \textbf{A solo activity} is performed by one person. Interaction between the actor with other people does not change the nature of this solo activity. E.g. driving, writing.
    \item \textbf{A group activity} is performed by at least two people interactively. The activity will not be the same without this interaction. E.g. wrestling, Tango dancing.
\end{itemize}

Based on these definitions, we can categorise activities by each of the above criteria separately. For example, a stretching exercise is a simple activity due to its requirement for proper form. It is a composite activity because it involves a series of different sub-activities. It is a solo activity because it can be practised alone.

\section{Unraveling the Causes of Inaccuracy}\label{sec: factors}
\begin{figure*}[htbp]
  \centering
  \includegraphics[width=0.5\linewidth]{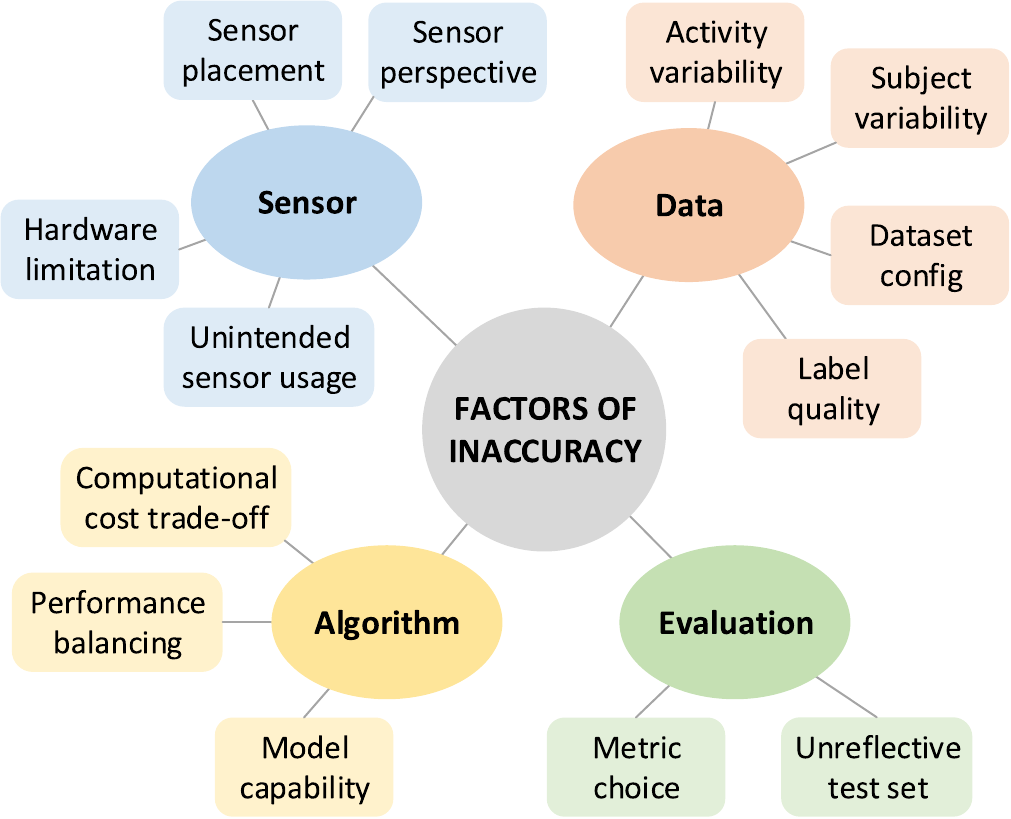}
  \caption{Factors of Inaccuracy in Complex HAR}
  \label{fig:factors}
\end{figure*}

This section systematises factors that impact the performance of complex HAR using deep learning models. As mentioned in \Cref{sec:intro}, HAR performance depends on the sensor, data, algorithm, and activities of interest, but only the first three can be controlled. The evaluation method also influences HAR accuracy indirectly. We divide all factors into these groups, summarised in \Cref{fig:factors}. Although some may not fit neatly into a single group, we make an effort to group them logically to aid in analysing and solving these issues. Many examples are given throughout this section to support our points, especially examples about fall detection because it is widely studied as a standout case of complex activities.

\subsection{Sensor}\label{subsec: sensor}
Data recording is the first step of any HAR system, all downstream tasks will be affected if the sensors do not work properly. This is why sensors must be prepared rigorously. We use the term \textit{sensor} to refer to all sensing devices, including wearable, camera, radar, etc.

\subsubsection{Sensor placement}\label{subsubsec: sensor placement}
Sensor placement must be considered carefully when developing a HAR system because it directly affects HAR's input data. Different sensor placements result in different distributions of data fed into the model.

For wearables, sensor positions on the body are very important. Sensors worn on positions with more movements like wrist or ankle introduce more noise into the deep learning model, while hip or waist is more suitable if we only need to capture whole body movements. Also, users do not always wear sensors with the same orientation, leading to inaccuracy if not considered in model building \cite{Shahzad2018}.

For ambient sensors like camera, radar, and infrared sensor, sensor position affects the observation range, thus it has some influence on HAR accuracy \cite{Bodor2005}. However, this is expected for ambient sensors and is often taken into account \cite{Ali2018}.

\subsubsection{Sensor perspective} \label{subsubsection: Sensor perspective}
In many situations, a single sensor cannot provide the deep model with sufficient information to learn the differences among activities, or the differences are too subtle for the model to spot.

Though accelerometer is the most commonly used inertial sensor, it has many limitations. For instance, wrist-worn accelerometers cannot recognise gestures like wrist turning \cite{KWOLEK2016}. The use of both accelerometer and gyroscope together could help reduce misclassification. Since both sensors can be built into a single device, it is not more intrusive than using one sensor. Nevertheless, even a wearable device with both an accelerometer and a gyroscope cannot observe environmental factors, thus it has a limited ability to recognise many activities that have similar inertia. Also, a sensor worn on a body position cannot capture motions of other parts of the body, which could be indicative of the subject's activities.

On the other hand, ambient sensors, including camera, sense everything in their ranges, thereby also producing more environmental noise in the data and consequently causing inaccuracy. Any pet or other people besides the targeted subject are also environmental factors and would introduce significant disturbance and noise to the system, as ambient sensors do not focus solely on the targeted subject \cite{Wangzhuo2020}. Note that ambient sensor is a broad term for many specific types of sensors, so they are influenced by different noise sources.

To name some examples: acoustic, floor pressure, and ultrasonic sensors in fall detection may produce false alarms when a large pet falls on the floor, but Doppler radar and vibration sensor may be able to tell the difference \cite{Singh2020}; sitting down quickly and jumping generate large acceleration signals like a fall would \cite{HAKIM201746}; lying down on the floor and sitting down suddenly are not easily distinguishable from falling only by viewing from an RGB camera \cite{WangXueyi2020}; activities with subtle motions like reading book, writing, and using laptop cause many errors for depth camera-based HAR \cite{Yang2017}; writing and drawing mostly resemble for many sensors. These resembling activities are likely to confuse HAR systems.

\subsubsection{Unintended sensor usage}
In the real world, sensing devices could be used in unintended ways, creating unusual signals that may act as disturbances to the model. \cite{Kangas2014} tested a fall detection wearable prototype in a real-world condition. The author observed that most reported false alarms were related to taking off or attaching the sensor. \cite{Chaudhuri2015} also tested a wearable device and their report showed that 16.9\% false alarms happened when the participant dropped the device, while device misuse and putting down the device accounted for 10.8\% of false alarms each. Non-wearable sensors are less likely to encounter these problems in usage since they do not have as much close interaction with users.

\subsubsection{Hardware limitation}
Sensor units of the same type may operate divergently, especially if they are not properly calibrated. As an example, gyroscope experiences time-varying zero shift \cite{Rastogi2021}, and the shift value differs for each device. This can result in disparities between training data and usage time data, or between datasets.

As HAR data are often time series, the sampling rate also contributes to the deep model's classification accuracy \cite{Yang2023}. Hence, systems using low-powered sensor devices have to deal with the trade-off between computational cost and accuracy. Additionally, network failures are a persistent challenge in sensor networks, leading to data loss and the drop of sampling rates below their intended levels.

\subsection{Data}
The divergence between training data and inference data is a major problem. It would be very beneficial to have real-world data for HAR model development. However, real-world data collection encounters various difficulties, including issues related to sensor intrusiveness, sensor usability, scheduling, experimental environment, expenses, annotations, and more \cite{Wang2019}. In the experiments conducted by \cite{Bourke2010}, a fall detection algorithm reached 100\% both sensitivity and specificity on a scripted dataset but gave false alarms in a real-world condition. Besides the disparity caused by sensor devices (\Cref{subsubsec: sensor placement}), this section discusses various factors contributing to data distribution divergence as below.

\subsubsection{Activity variability}
A complex activity can be performed in varied ways (i.e. intra-class variability). This variation is serious in the real world, where innumerable possibilities exist but cannot be foreseen when building HAR systems. Also, there could be different activities performed similarly (i.e. inter-class similarity). For example, fall is a rare event but occurs in diverse ways \cite{Debard2011}, thus collecting a good dataset for training is very challenging. Due to this diversity, many non-fall activities share similar characteristics with falls, leading to false alarms if they are not included in the training set \cite{Aziz2017}.

Some activities are very difficult to simulate in controlled lab environments. This is especially true for activities that are supposed to be unexpected, such as falling, fainting, loss of balance, muscle cramp, inappropriate behaviours (in surveillance applications), etc. Specifically, when collecting data for fall detection, the subject would expect to fall and be ready for it, which does not happen in real falls. The body would instinctively prepare itself to minimise the impact \cite{Nguyen2023}.

\subsubsection{Subject variability}
Training and inference data also differ because actual users have different characteristics from dataset participants, which leads to misclassification. It is a very common and good practice to split datasets into a training set and a test set by subjects, i.e. cross-subject evaluation. However, this is still not sufficient because most datasets involved only young and healthy subjects for the obvious reason that old or ill people must not risk their health and safety for dataset collection. Some other examples of public simulated datasets are KFall (subjects age 24.9\textpm3.7) \cite{yu2021_kfall}, UP-Fall (subject age 18-24) \cite{Villasenor2019}. The age groups of subjects were mostly 20s, whereas, in real-world applications, user age may vary greatly. SisFall \cite{Sucerquia2017_sisfall} is a fall detection dataset which has 23 young adults and 15 older participants (age 60-75), in which one older person could simulate falls.

\subsubsection{Dataset configuration}
Datasets are configured differently when recording. Therefore, if we build a HAR model using many data sources, it is crucial that all data be configured and processed in the same way to ensure consistency and prevent biases that could negatively impact the model's performance. A particular case is that each sensor unit may have a different measuring range. For example, the accelerometer used in KFall measured acceleration signal between the range \textpm16g, while the range is \textpm8g for FallAllD \cite{saleh2020_fallalld}. Also, each dataset may apply its own data pre-processing steps. To illustrate, accelerometer data in FallAllD is raw, whereas HAPT \cite{jorge2015_hapt} applied noise filters to accelerometer and gyroscope signals. Acceleration signals of a fall after smoothing may lose some magnitude and become more similar to raw signals of non-fall activities. According to \cite{Sucerquia2017_sisfall}, applying a low-pass filter on accelerometer and gyroscope data do not always improve fall detection accuracy.

\subsubsection{Label quality}
Manual label accuracy is a major challenge due to inherent subjectivity and inconsistency in human annotation \cite{Rosenthal2010,Cruciani2018}. Human annotators may interpret activities differently, leading to discrepancies in labelling. Moreover, clear starting and ending points of some complex activities cannot be straightforwardly determined in a time series. For example, discerning the exact moment when the person starts cooking dinner, as opposed to merely preparing ingredients, can be difficult due to the gradual and overlapping nature of these activities. These issues hinder the reliability of labelled datasets, impacting the accuracy of HAR models.

In the specific case of anomaly detection, the model is trained to recognise normal activities, and everything else is considered an anomaly. Adversely, it is important to establish clear definitions of normal activities. If there is a lack of data on normal activities, such approaches may lead to an abundance of false alarms \cite{KHAN201712}.

\subsection{Algorithm}
Given the same training data, different algorithms exhibit varying performance.

\subsubsection{Computational cost trade-off}
Sometimes HAR model's accuracy has to be reduced in exchange for a lower computational cost \cite{Wang2022}, e.g. resource usage in edge devices, and runtime in large-scale systems. In fact, the trade-off between accuracy and computational cost applies to every task and not only HAR. However, this problem is especially relevant for HAR applications where real-time processing and continuous operation are often crucial. Furthermore, resource-limited devices like wearables play an important role in many HAR systems. These devices typically have limited processing power and memory, making it challenging to run computationally intensive models. Powerful models based on deep learning can extract high-level features from activity data, but their resource requirements are a hurdle for resource-constrained devices.

\subsubsection{Performance balancing}
It is infeasible to achieve perfect accuracy in real-world scenarios. Consequently, a trade-off must be carefully considered between precision and recall (or specificity and sensitivity) \cite{Palmerini2020}, which both are crucial metrics encapsulating the model's performance. The choice between prioritising which metric depends on the specific application's requirements and the relative importance of avoiding false positives and false negatives. For example, specificity is generally more important than sensitivity for a fall detector because if it misses 1 in every 10 falls, it is a bad detector, but if it produces 1 false alarm every 10 guesses, it is unusable.

In vision-based HAR, there is an additional precision-recall trade-off when the system is tasked with identifying the subject in images \cite{Padilla2020}. This applies to systems utilising either an object detection or a pose estimation model. A low precision creates much noise for the downstream task (activity classification), while a low recall causes missing data for it.

\subsubsection{Model capability}
Modern deep learning-based HAR models are very powerful but still have much trouble recognising complex activities. In \Cref{subsubsection: Sensor perspective} we have discussed the problem of resembling activities. Nonetheless, it could also be considered an algorithm-related issue, given the subtle differences among such activities that classifiers often struggle to recognise. Many resembling activities are expected in daily life. Adding data of such activities to the training set could improve the overall performance. However, it is infeasible and impractical to define and gather data and annotations for every possible scenario. It is desirable that classifiers can find the differences among them, even with ambiguous clues and limited training scenarios.

The capability of a deep learning model lies in not only its network architecture but also its training strategy. A suitable optimisation function, learning rate schedule, and loss function contribute to HAR performance. For instance, with the same training dataset and deep network architecture, a self-supervised model based on contrastive learning can outperform a conventional supervised learning model \cite{Deldari2022,Jain2022}. Furthermore, adding handcrafted features calculated from the input data itself can improve accuracy \cite{Koutrintzes2022}. This suggests that HAR models can still be improved to learn more from the same amount of data, either labelled or unlabelled.

\subsection{Evaluation}
Evaluation indirectly influences the performance of HAR in real applications. HAR systems must be tested thoroughly before they can be used, otherwise, they may perform well in the tests but poorly in actual usage.

\subsubsection{Unreflective test set}
Data-related questions are perennial challenges, affecting both training and test data. An effective test set should accurately reflect the model's performance in real-world usage. An overly simplistic test is not a reliable tool for model comparison, as it may yield scores that do not differentiate between inferior and superior methods.

As for the test set's class label distribution, imbalance tolerance metrics like F1-score and AUC can be used. However, these metrics may not always meet the application's needs. Thus, it is better to have a test set with class label distribution that mirrors the real-world occurrence rates of target activities. In the case of simulated fall datasets, there is a large number of falls for evaluation, whereas falls are rare in real life. This can affect HAR performance evaluation, enlarging the false positive rate \cite{Xefteris2021}. As a result, metrics calculated on these datasets may not reflect the model's performance. For example, the precision score is calculated as: $precision = TP/(TP+FP)$. In a dataset with a lot of fall events, $TP$ becomes higher, and precision increases with it. By contrast, in a realistic test set with only a few fall events and a large number of negatives, $FP$ becomes higher and precision decreases.

\subsubsection{Metric choice}
Each evaluation metric shows an aspect of model performance. Along with a good test set, the choice of indicative metrics is essential for assessment and comparison. In a fall detection paper \cite{Palmerini2020}, an ostensibly optimal AUC score of 0.996 translates to 88\% sensitivity with more than one false alarm per hour. This implies AUC is not a suitable metric in this case, and an algorithm with very high scores on a test dataset may still perform poorly in actual usage.

\begin{table*}[htbp]
\centering
\caption{Main focuses of several deep learning-related HAR papers}\label{tab: summary of methods}
\begin{tabular}{@{}lp{0.17\linewidth}p{0.6\linewidth}@{}}
\toprule
\textbf{Paper} & \textbf{Targets} & \textbf{Description} \\ \midrule

\makecell[t]{\cite{Abedin2021}\\2021} & Algorithm (capability), Data (activity variability), Sensor (perspective) & A framework including (1) a multimodal channel attention encoder, (2) a temporal attention encoder, (3) a centre-loss to reduce intra-class variability, (4) mixup augmentation to improve generalisation. \\\midrule

\makecell[t]{\cite{Meng2021}\\2021} & Algorithm (computational cost) & This method works with video   data. A policy network decides whether to keep or skip channels from a frame, or reuse the same channel from the previous frame. This reduces the overall computational cost. \\\midrule

\makecell[t]{\cite{Islam2021}\\2021} & Sensor (perspective) & Main steps: (1) extracting unimodal features, (2) retrieving unimodal features using multimodal context, (3) capturing multimodal relationships, generating complementary multimodal features, (4) classification using multimodal features. \\\midrule

\makecell[t]{\cite{Haresamudram2021}\\2021} & Data (variability) & Pre-training on unlabelled data using the Contrastive Predictive Coding (CPC) framework, fine-tuning on labelled data for classification.
\\\midrule

\makecell[t]{\cite{Ambriz2022}\\2022} & Algorithm (capability) & A deep network for video data   consisting of a 3D CNN and a Convolutional LSTM network. \\\midrule

\makecell[t]{\cite{Tang2022}\\2022} & Algorithm (capability), Sensor (perspective) & Given multimodal data with 3 dimensions, i.e. [sensor$\times$time$\times$channel], this method involves 3 attention branches, learning the interaction within each of 3 pairs of dimensions, respectively. Three branches are aggregated for classification. \\\midrule

\makecell[t]{\cite{Tang2022_multiscale}\\2022} & Algorithm (capability) & A Hierarchical-split CNN learns multiscale features using multiple network branches with different receptive fields. All scales are combined for classification. \\\midrule

\makecell[t]{\cite{Jain2022}\\2022} & Data (variability), Sensor (perspective) & Pre-training with contrastive learning among multiple sensors, fine-tuning on one sensor for classification. Positive and negative samples for contrastive learning are chosen based on similarity and time steps.\\\midrule

\makecell[t]{\cite{Haresamudram2023}\\2023} & Algorithm (capability),   Data (variability) & New architectures for 3 components of the CPC framework, i.e. encoder, aggregator, and future timestep prediction. \\\midrule

\makecell[t]{\cite{Hu2023}\\2023} & Data (subject variability) & Domain adaptation for unlabelled data of a new subject. The model learns sample weights, samples with large weights are trained more to remove patterns separating the 2 domains.\\ \bottomrule

\end{tabular}
\end{table*}

\Cref{tab: summary of methods} shows some of the most cited deep learning-related HAR papers from reputable publishers in recent years. The table outlines the aspects they targeted with respect to the taxonomy above and their proposed approaches.

\section{Towards Robust Complex Activity Recognition}\label{sec: towards}
In this section, we delve into the current and potential future research areas aimed at addressing the above factors, highlighting both challenges and opportunities. Studies have proposed multi-branch neural network architectures to capture multiple aspects of the data (\Cref{tab: summary of methods}). Besides that, many researchers are focusing on other areas as below.

\subsection{Domain generalisation}
Data augmentation is a popular technique for improving domain generalisation. Though sensor device placement can be adjusted for a more informative view and less noise, sometimes it is restricted to application requirements and user preference. Data augmentation can simulate possible scenarios not present in the dataset, including sensor placements. For instance, a 3D rotation operation can simulate different orientations of wearables. \cite{Cormier2024} proposed skeleton augmentation techniques (e.g. occlusion, interpolation, keypoint swapping) that simulate errors in human pose estimation to train a more robust downstream HAR model. There are also studies generating new training samples using GAN \cite{Hoelzemann2021} and genetic algorithm \cite{Nida2022}.

There are methods aiming at training domain-invariant models, i.e. producing the same latent distribution regardless of input domains. For example, to achieve this purpose, \cite{lu2024} uses Gradient Reversal Layers to make the model not able to distinguish different domains.

\subsection{Multimodal approaches} \label{subsec: multimodal approaches}
Sensor fusion involves combining data from multiple modalities to create features, allowing the model to learn from this combination. Besides that, recent studies have also proposed models learning the correlations between modalities when fusing them, using contrastive learning \cite{Deldari2022}
. However, in many situations, the type and number of sensors are limited due to high costs or privacy problems. There are several approaches to take advantage of multimodal data for training while using only one sensor for inference, such as co-learning (or mutual learning) and contrastive learning.

Co-learning fosters knowledge transfer among modalities, allowing for independent inference from each modality. Co-learning can handle unlabelled or noisy data to improve robustness \cite{Rahate2022}. As an example of this approach, \cite{Qu2023} trains a main model and an auxiliary model together, exploiting both labelled and unlabelled data with semi-supervised co-learning.

While co-learning is often used in a semi-supervised setting, contrastive learning can be used in a self-supervised setting where the unlabelled data do not necessarily contain the same human activities as the labelled data. This is because it learns the correlation among modalities instead of learning to generate pseudo labels. For instance, ColloSSL \cite{Jain2022} learns the correlation between the main modality and others with similar data distributions, whereas Virtual Fusion \cite{nguyen2023virtual} learns that between every pair combination. Unlike the aforementioned contrastive learning methods that fuse all modalities, these methods only use one or a subset of modalities for inference.

\subsection{Semi-supervised learning and self-supervised learning}
One straightforward way to address the disparity between training data and real-world data is to collect a large dataset in the environments where HAR system deployment is intended. However, human data collection and annotation pose significant challenges due to privacy and intrusiveness concerns. Annotating data in non-human-readable formats, such as inertial signals, requires even more effort. On the other hand, collecting unlabelled data is comparatively more feasible. Semi-supervised learning and self-supervised learning approaches are aimed to leverage this abundant data source.

Besides the methods mentioned in \Cref{subsec: multimodal approaches}, there are many other semi-supervised and self-supervised methods proposed for unimodal HAR. The semi-supervised learning method in \cite{Bi2023} creates pseudo labels by applying temporal ensemble on unlabelled data. The labels deduced through temporal ensemble are utilised as training targets for the unlabelled instances. \cite{Taghanaki2021} proposed a self-supervised learning architecture in which the self-learning task is to recreate a missing piece in the raw data. JDS-TL \cite{li2022} is a semi-supervised method transferring knowledge from a labelled source dataset to a sparsely labelled target dataset using an adversarial loss to align the distributions between the two domains.

\subsection{Computational cost optimisation}
Researchers have been trying various ways to reduce deep learning models' computational cost while minimising accuracy loss. \cite{Zhou2022} designed a lightweight deep model for HAR based on convolutional and transformer architectures. \cite{Xu2023} developed a distillation method to transfer knowledge from a large teacher model to a smaller student model while taking into account the bias introduced by the teacher. There are also papers proposing systems with a tunable trade-off between computational cost and accuracy \cite{Wang2022, Yang2023}.

For deployment, there are many tools for optimisation such as Nvidia TensorRT, ONNX Runtime, Pytorch Mobile, and Tensorflow Lite. They support many optimisation techniques, including using low-precision number formats (i.e. quantisation), removing redundant operations within the model, combining network layers that can be executed together, and some hardware-specific optimisations.

\subsection{Fair and transparent evaluation}
Comparing new methods with prior papers is a common way to evaluate their quality. \cite{Zhang2022} compiled a list of 23 major HAR datasets based on the number of citations per year. Among these datasets, WISDM, UCI-HAR, MHEALTH, Opportunity, and PAMAP2 are often used by researchers as benchmarks for comparison. However, Opportunity is the only one with complex activities used for model evaluation.

To ensure fairness, all methods in the comparison must use the exact same test dataset. Upon examining the 23 datasets above, we discovered that 9 of them include complex activities, but only 1 simple activity dataset (UCI-HAR) facilitates fair comparison by providing a clear train-test split and segmented data. If time-series data is not pre-segmented into windows or sessions, serving as individual samples in the test set, different authors may segment it in varying ways, leading to a different test set for each paper using the dataset. Also, each paper may use dissimilar metrics to serve its purposes. Thus, quick comparisons cannot be made. An unnormalised confusion matrix would be useful to include as most common metrics can be calculated from it.

Finally, the lack of transparency in many papers obstructs their reproducibility. A review paper \cite{Islam2022} showed that a lot of papers in their analysis did not include detailed method configurations, and none provided their implementation. If an experiment cannot be reproduced using the information in its paper, questions will be raised about the reliability of the work.

\section{Conclusion}
HAR has been studied for several decades, witnessing remarkable advancements. Despite that, modern deep learning methods for HAR still grapple with the recognition of complex activities. This paper provides a comprehensive systematisation of factors contributing to the low accuracy of complex HAR using deep learning, divided into four main categories, namely sensor, data, algorithm, and evaluation. We also review and discuss a range of potential methods for further research. New studies tend to scale deep network architectures in width (multiple branches). Researchers are also increasingly focusing on exploring more model training strategies. The potential of vast unlabelled data and multimodal correlations can be further exploited. Meanwhile, more work could also be done to promote fair and transparent model evaluation, which aids in the discovery of novel ideas.

\bibliographystyle{IEEEtran}
\bibliography{ref}

\end{document}